\documentclass[pre,twocolumn, amsmath, amssymb, superscriptaddress]{revtex4}

\usepackage{graphicx}
\usepackage{dcolumn}
\usepackage{bm}
\usepackage{braket}
\usepackage{xcolor}

\usepackage{amsmath}

\newcommand{\beq}{\begin{equation}}
\newcommand{\eeq}{\end{equation}}

\begin{document}

\title{High throughput interactome determination via sulfur anomalous scattering}


\author{Mattia Miotto\footnote{Corresponding author: mattia.miotto@iit.it}}
\affiliation{Center for Life Nano \& Neuro Science, Istituto Italiano di Tecnologia, Viale Regina Elena 291,  00161, Rome, Italy}

\author{Edoardo Milanetti}
\affiliation{Department of Physics, Sapienza University, Piazzale Aldo Moro 5, 00185, Rome, Italy}
\affiliation{Center for Life Nano \& Neuro Science, Istituto Italiano di Tecnologia, Viale Regina Elena 291,  00161, Rome, Italy}

\author{Riccardo Mincigrucci}
\affiliation{Elettra-Sincrotrone Trieste S.C.p.A. di interesse nazionale Strada Statale 14 - km {163.5} in AREA Science Park 34149 Basovizza, Trieste ITALY}

\author{Claudio Masciovecchio}
\affiliation{Elettra-Sincrotrone Trieste S.C.p.A. di interesse nazionale Strada Statale 14 - km {163.5} in AREA Science Park 34149 Basovizza, Trieste ITALY}

\author{Giancarlo Ruocco}
\affiliation{Center for Life Nano \& Neuro Science, Istituto Italiano di Tecnologia, Viale Regina Elena 291,  00161, Rome, Italy}
\affiliation{Department of Physics, Sapienza University, Piazzale Aldo Moro 5, 00185, Rome, Italy}

\begin{abstract} 
We propose a novel approach to detect the binding between proteins making use of the anomalous diffraction of natively present heavy elements inside the molecule 3D structure. In particular, we suggest considering sulfur atoms contained in protein structures at lower percentages than the other atomic species. Here, we run an extensive preliminary investigation to probe both the feasibility and the range of usage of the proposed protocol. In particular, we (i) analytically and numerically show that the diffraction patterns produced by the anomalous scattering of the sulfur atoms in a given direction depend additively on the relative distances between all couples of sulfur atoms. Thus the differences in the patterns produced by bound proteins with respect to their non-bonded states can be exploited to rapidly assess protein complex formation. Next, we (ii) carried out analyses on the abundances of sulfurs in the different proteomes and molecular dynamics simulations on a representative set of protein structures to probe the typical motion of sulfur atoms.  Finally, we (iii) suggest a possible experimental procedure to detect protein-protein binding. Overall, the completely label-free and rapid method we propose may be readily extended to probe interactions on a large scale even between other biological molecules, thus paving the way to the development of a novel field of research based on a synchrotron light source.
\end{abstract}

\maketitle

\section{Introduction}

Protein-protein interactions play a crucial role in various biological processes, including signal transduction, enzymatic regulation, and molecular recognition~\cite{bonetta2010interactome}. Understanding the mechanisms and dynamics of these interactions is essential for elucidating cellular processes and developing therapeutic interventions. 
 Given the importance of the knowledge of protein-protein interactions (PPI), several experimental techniques have been developed in the last decades based on biochemical and/or biophysical methods. The former include co-immunoprecipitation, bimolecular fluorescence complementation (BiFC)~\cite{Hu2002}, phage display ~\cite{sidhu2003exploring}, tandem affinity purification (TAP) ~\cite{rohila2006protein, bailey2012identification, goodfellow2014detection}, and Proximity ligation assay (PLA) ~\cite{soderberg2006direct, hegazy2020proximity}. Techniques based on biophysical processes comprise surface plasmon resonance (SPR), dual polarisation interferometry (DPI), flow-induced dispersion analysis (FIDA), Fluorescence resonance energy transfer (FRET), and Bio-layer interferometry (BLI)~\cite{Rich2007, Piacentini2022}.

\begin{figure*}[t]
\centering
\includegraphics[width = \textwidth]{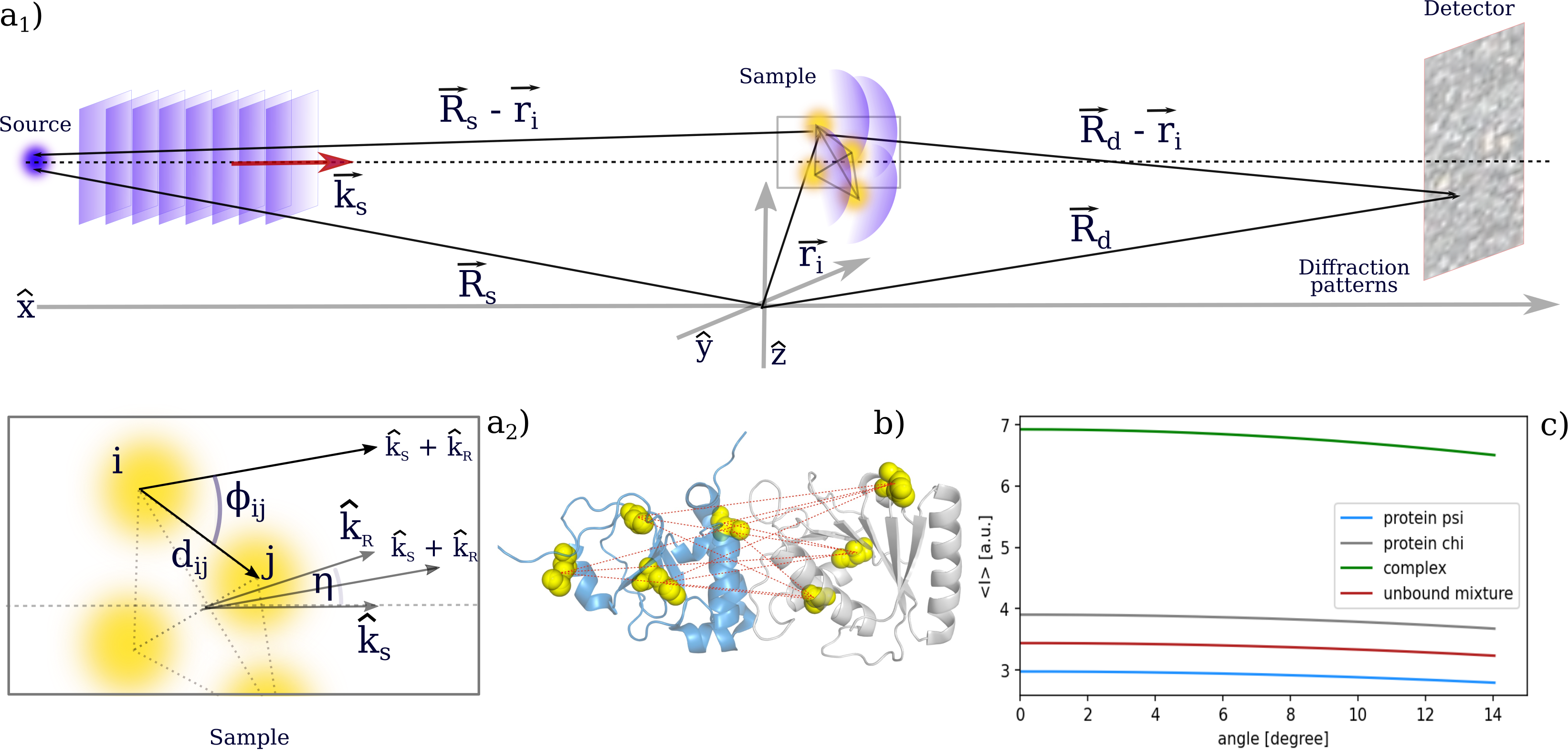}
\caption{\textbf{Scheme of the scattering process and case-of-study results.} 
\textbf{a1)} Sketch of the scattering process. \textbf{a2)} Zoom on the sample with a schematic representation of the key geometrical quantities. \textbf{b)}  Cartoon representation of the simulated complex (PDB: 1EM8) with the position of the sulfur atoms highlighted in yellow.  In particular, proteins A and B comprise 3 and 4 sulfur atoms, respectively. \textbf{c)} Radial intensity of the signal registered on the detector as a function of the observation angle. }
\label{fig::1}
\end{figure*}

\begin{figure*}[]
\centering
\includegraphics[width = \textwidth]{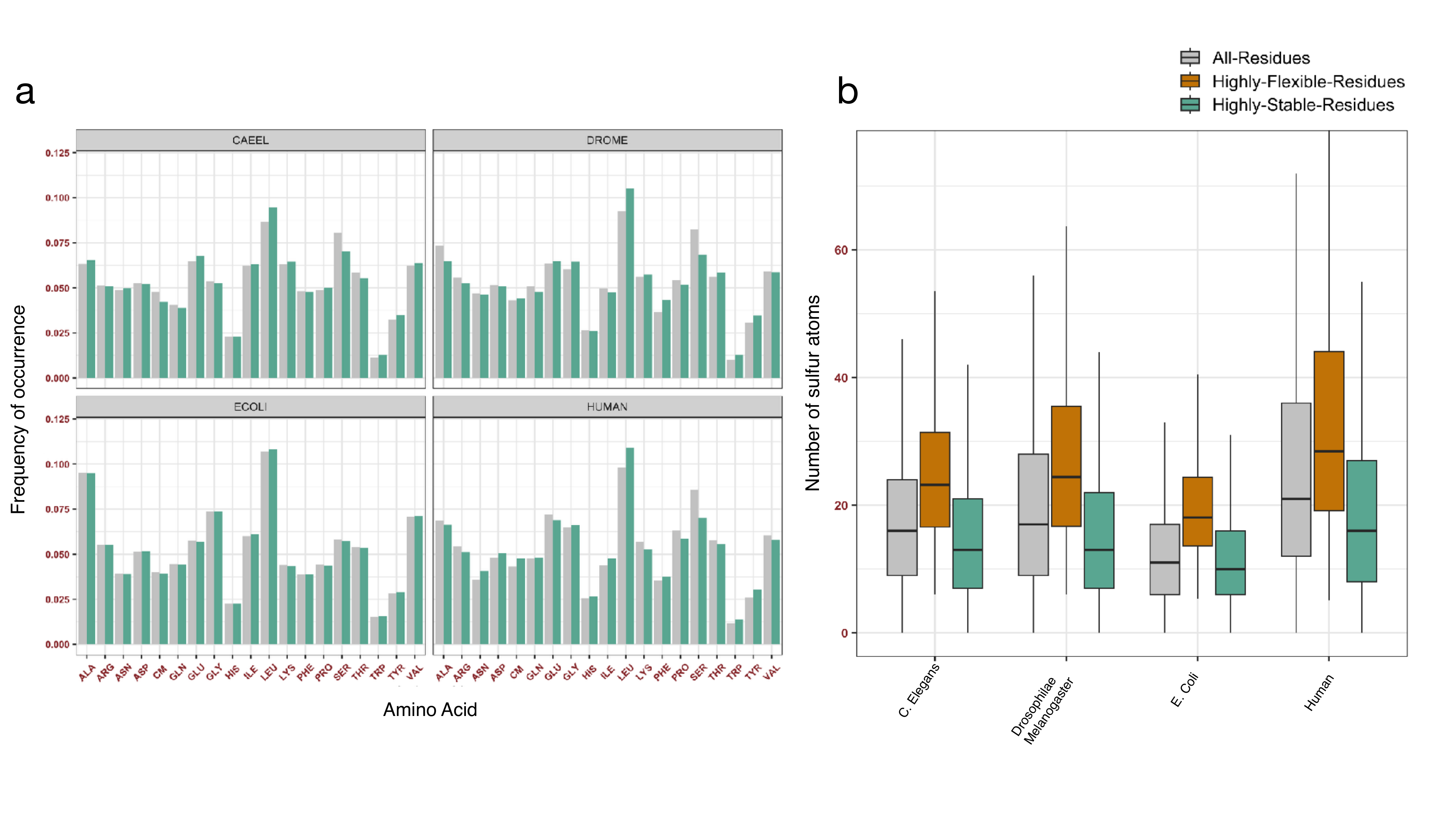}
\caption{\textbf{Abundance of sulfur atoms in protein structures.} \textbf{a)} Frequency of occurrence of the twenty natural amino acids in the proteome of four representative organisms, i.e. \textit{C. Elegans}, \textit{Drosophilae Melanogaster}, \textit{E. Coli}, and \textit{Human}. Grey bars refer to all protein residues, while green ones correspond to residues predicted to be in structured, highly stable regions, according to the AlphaFold2 score.  \textbf{b)}  Box plot representation of the distributions of the number of sulfur atoms per protein found in the proteome of the four representative organisms.  Grey bars refer to all protein residues, while green (respectively orange) ones correspond to residues predicted to be in structured, highly stable (resp. flexible) regions, according to the AlphaFold2 score. }
\label{fig::2}
\end{figure*}

Each of them offers different insights into the nature and characteristics of the interactions~\cite{Miura2018}. However, up to now, a label-free technique able to rapidly assess whether and where two proteins bind is not available yet. To get detailed information on the protein-protein complex structure, we rely on X-ray, NMR, or cryo-EM experiments able to provide the spatial configuration of the complex up to a certain resolution~\cite{wang2017cryo}. 
Unfortunately, these experimental methodologies are time-consuming and heavily depend on the kinds of protein complexes and the experimental conditions~\cite{kang2019applications, oviedo2019fast}. For this reason, the development of new, fast, structure-driven experimental techniques to assess protein-protein binding is of paramount importance, especially in the era of artificial intelligence aimed at predicting the three-dimensional conformations of protein structures (see for instance the recently developed methods such as AlphaFold2~\cite{jumper2021highly} and RoseTTA fold~\cite{baek2021accurate}). Experiments will be requested to test huge amounts of computationally predicted interactions as well as to increase the training, and therefore the performance, of all those emerging predictive methods based on machine learning data~\cite{jumper2021highly}.

In this perspective article, we suggest using anomalous scattering from native sulfur atoms to rapidly assess whether or not a couple of proteins form stable complexes. Moreover, depending on the number of native sulfur atoms, we expect that by combining the information coming from interference patterns with the knowledge of the two unbounded protein structures, it will be possible to obtain insights into the binding region.

In this respect, sulfur anomalous scattering has been already applied in the contest of protein structure determination. It resulted in being a valuable tool in X-ray crystallography for studying protein structure and addressing phase ambiguity or dephasing problems during structure determination~\cite{DAUTER199983}.  In X-ray crystallography, knowing the phases of diffracted X-rays is crucial for determining the electron density of a protein crystal, which subsequently reveals the protein's atomic structure.  However, dephasing problems can arise due to the phase ambiguity inherent in the X-ray diffraction pattern, especially when the resolution of the data is limited or when the crystal lacks heavy atoms.     Sulfur atoms in proteins exhibit anomalous scattering, which arises from X-ray interactions with sulfur's electron configuration.   This phenomenon leads to differences in X-ray scattering between wavelengths, providing additional experimental data for solving phase ambiguity and improving electron density maps.
In particular, Multi-wavelength Anomalous Dispersion (MAD) is a technique used in X-ray crystallography to overcome phase ambiguity problems~\cite{Guss1988}.   By collecting X-ray diffraction data at multiple wavelengths around the absorption edge of sulfur (in the range of $1.77 - 2.07$ $\AA$), the anomalous differences between the wavelengths provide additional phase information, allowing for more accurate structure determination. Similarly, the  Single-wavelength Anomalous Dispersion (SAD) approach involves collecting X-ray diffraction data at a single wavelength near the absorption edge of either native atoms, atoms used in the crystallographic process, or atoms inserted \textit{ad hoc} in the structure~\cite{Rose2016, Panneerselvam2017}.   Although SAD requires fewer experimental data compared to MAD, it still offers enough anomalous signal from sulfur to deduce phase information and resolve the protein structure~\cite{Rose2016}.

Here, we propose to use anomalous scattering to detect protein-protein interactions without the need for a crystallographic structure.  Our intuition is to compare the sum of the signals of the photons scattered by native sulfur atoms for ultra-diluted solutions of each protein independently, with that of the solution containing both proteins: the sum of the signal produced by unbound proteins is predicted to differ from the signal produced by the two protein when bound together.

\section{Results}

The novel experiment we propose aims at the fast determination of protein complex formation by exploiting the differences in the intensity of the diffraction patterns produced by the anomalous scattering of the sulfur atoms of the two potentially interacting proteins: not interacting proteins will produce an intensity made by the superposition of the patterns produced by the two proteins found alone in solution, which will differ from the pattern produced by the proteins forming a complex. 

To get analytical insights on this crucial aspect, we consider the model setup described in Figure~\ref{fig::1}a1-a2, depicting the diffusion process of a coherent excitation source on a sample. The coherent excitation source, that can be thought of as a plane wave $\phi_S= A_s e^{-i \vec {k_s} \cdot \vec{r}}$, where $|k_s| = 2\pi/\lambda$ and $\lambda$ is in the range around the sulfur K-edge ($\sim 0.5$ nm). Interaction between the source and the proteins sulfur atoms around their K-edge threshold produces a scattering process, that can be approximated by spherical waves originating from each sulfur atom. 
The electromagnetic field at position $\vec{R}_d=(x_D, y,z)$ on the detector is given by the scattered light of the $N$ protein' sulfur atoms is given by:

\begin{equation}
    \psi = \sum_i^N \psi_i
\end{equation}

where the field produced by the i-th atom has the form 

\begin{equation}
    \psi_i = \frac{A_i~e^{-i |\vec{k^i_f}|  | \vec{R}_d - \vec r_i|}}{|\vec{R}_d - \vec r_i|} ~e^{-i \vec{k}_s \cdot (\vec{R}_s - \vec{r}_i)} 
\end{equation}

where $\psi_i$ is given by the product of a spherical wave of amplitude $A_i$ and wave vector $|\vec{k}^i_f| = \frac{2\pi}{\lambda}$
 with a phase term depending on the distance between the source and the i-th sulfur atom.
Assuming that one can measure one protein/complex at a time, the intensity at the detector will depend on the number and disposition of the sulfur atoms in the system:

\begin{multline}
\label{eq:I}
  I(\vec{R}_d) = | \psi |^2  = \\ =  \frac{A^2}{|R_d|^2} \sum_{i,j}^N  e^{-i \left( |\vec{k}^i_f|| \vec{R}_d - \vec r_i| - |\vec{k}^j_f||\vec{R}_d - \vec r_j|\right)} ~e^{ -i \vec{k}_s \cdot (\vec{r}_j - \vec{r}_i)} 
\end{multline}

where we assumed that spherical waves possess the same amplitude and that the distance of the detector between the sample and the detector is such that $|\vec{R}_d| >> |\vec{r}_i|$ for $i = 1,..,N$. In the latter approximation regime, we can consider that the direction of propagation of the spherical waves is parallel, i.e. $\hat{k}^i_f \sim \hat{k}^j_f$, so that $\left( |\vec{k}^i_f|| \vec{R}_d - \vec r_i| - |\vec{k}^j_f||\vec{R}_d - \vec r_j|\right) \sim \frac{2\pi}{\lambda} \hat{k}_R\cdot (\vec{r}_j - \vec{r_i})$, where the versor $\hat{k}_R$ has the same direction of $\vec{R}_d$.  In this regime, Eq.~\ref{eq:I} reduces to:

\begin{multline}
\label{eq:Ired}
  I(\vec{R}_d) =  \frac{A^2}{|R_d|^2} \sum_{i,j}^N  e^{-i  \vec{q} \cdot \vec{d}_{ij}} = \\ = 
  \frac{A^2}{|R_d|^2} \left( N + 2\sum_{i<j}^{N} \cos{\left( \frac{4\pi}{\lambda}  ~\cos(\eta/2)~|d_{ij}|~\cos(\phi_{ij}) \right)} \right)
\end{multline}

with  $d_{ij} = |\vec{r}_i - \vec{r}_j|$, $\vec{q} = \vec{k}_R + \vec{k}_S$, $|\vec{q}| = \frac{2\pi}{\lambda}~2\cos(\eta/2)$, $\eta$ the convex angle between the versors $\hat{k}_R$ and $\hat{k}_S = \hat x$,  and $\phi_{ij}$ the convex angle between the $\hat{q}$ versor and the vector $\vec{d}_{ij} = (\vec{r}_i - \vec{r}_j)$.

Notably, the intensity depends on the distances between all couples of sulfur atoms in the system and on the orientation of the protein/complex. As measurements will be performed on proteins in suspension, the orientation of the protein/complex will be random. To remove the effect of the orientation, one can compute an average over different acquisitions, each being associated with an orientation uniformly distributed in the unit sphere. The outcome of the measure will be given by:

\begin{equation}
<I> = \int_0^{2\pi} d\theta \int_0^{\pi} d\phi I(\theta, \phi) \frac{\sin(\phi)}{4\pi} 
\end{equation}

where $\frac{\sin(\phi)}{4\pi}$ is the uniform distribution of sampling an orientation described by a couple of angles $(\theta, \phi)$ around the observation axis, $\hat{q}$.

After some straightforward calculations, one obtains 

\begin{multline}
\label{eq:Imean}
  <I> = \frac{A^2}{|R_d|^2} \left( N + 2\sum_{i<j}^{N} \frac{ \sin{\left( 4\pi \frac{|d_{ij}|}{\lambda}~\cos(\eta/2)\right)} }{4\pi \frac{|d_{ij}|}{\lambda}~\cos(\eta/2)} \right)
\end{multline}

where we implied that each couple of sulfur atoms can be found in all possible orientations with uniform distribution independently from the disposition of the other couples, i.e. we neglected correlations between the couples of sulfur atoms.   
As one can see, the final expression depends on the relative distances between sulfur atoms and the observational angle. 

To check the model, we ran a numerical simulation of an ideal, simplistic experiment outcome to test whether the disposition of the sulfur atoms in two interacting proteins produces a diffraction pattern distinct from the pattern produced by the two proteins alone in solution. 
 In Figure~\ref{fig::1}b-c, we consider a system of sulfur atoms mimicking that of the chi and psi subunit heterodimer from DNA polymerase III (pdb:1EM8). The two proteins contain three and four sulfur atoms, respectively (Figure~\ref{fig::1}b). Evaluating the outcome of a scattering process from a system composed by (i) just protein chi, (ii) just protein psi, (iii) the two proteins not bounded, and (iv) the complex, we obtained the signal in Figure~\ref{fig::1}c. It can be seen that the outcomes in the presence or absence of binding are distinguishable. 
 
To verify the range of applicability of the proposed technique, we performed a set of analyses, evaluating at first the abundance and distribution of amino acids containing sulfur atoms (i.e. methionine and cysteine) in the proteomes of different organisms which are usually used in protein-protein interaction investigations to determine the average number of sulfur atoms; next,  we analyzed the motion of sulfur atoms with respect to the motion of the whole protein structures, focusing on the relative distances between couples of sulfur atoms found in the structures of a representative set of twenty proteins.   Finally,  we propose an experimental apparatus and protocol to actually measure protein-protein interactions.

\begin{figure*}[t]
\centering
\includegraphics[width = \textwidth]{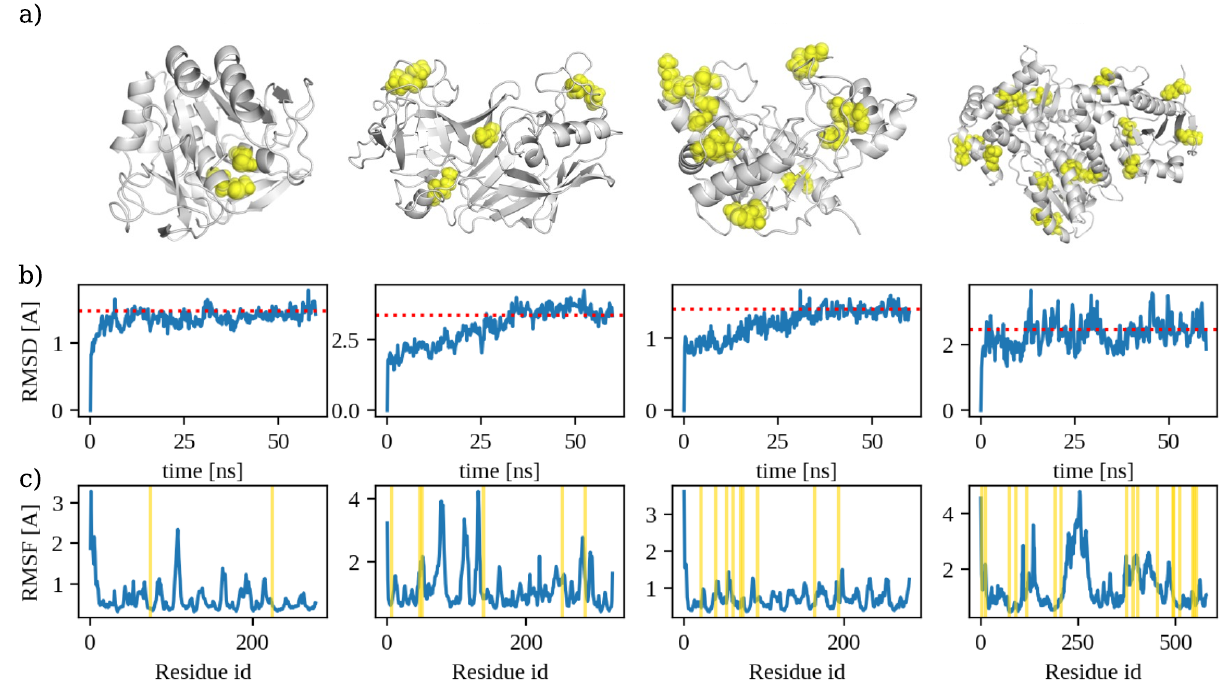}
\caption{\textbf{Analysis of Molecular Dynamics simulations.} 
\textbf{a)} Cartoon representation of four proteins from the selected dataset containing different numbers of sulfur atoms (in yellow).  \textbf{b)}  Root mean squared deviation (RMSD) as a function of the simulation time. The red horizontal line marks the equilibrium value.  \textbf{c)} Root mean squared fluctuations (RMSF) of each residue of the four chosen proteins. Yellow vertical lines mark the RMSF of the residues containing the sulfur atoms.}
\label{fig::3}
\end{figure*}

\subsection{Analysis on sulfur abundances in different organism proteomes}

To begin with, we performed a statistical analysis of the number of sulfur atoms in protein structures across different organisms. In particular, we select all sequences of the proteome of \textit{C. Elegans}, \textit{Drosophilae Melanogaster}, \textit{E. Coli}, and \textit{Human}, to study the distribution of the number of sulfur atoms per protein, operationally counting the number of methionine and cysteine, which are the only two natural amino acids containing a sulfur atom. 

Figure~\ref{fig::2}a reports the frequency of occurrence of the twenty natural amino acids. Grey bars refer to all protein residues, while green ones correspond to residues predicted to be in structured, highly stable regions, according to the AlphaFold2 score.  
AlphaFold2 produces a per-residue estimate of its confidence, which is called pLDDT, on a scale from 0 - 100. This confidence measure assigns to each residue a reliability value of the corresponding position prediction. The higher the pLDDT score the greater the reliability of the prediction. The authors of the AlphaFold2 algorithm demonstrated that pLDDT $< 50$ is a reasonably strong predictor of disorder, thus suggesting that this region is unstructured in physiological conditions~\cite{jumper2021highly}.

In figure~\ref{fig::2}b, the box plot representation of the distributions of the number of sulfur atoms per protein found in the proteome of the four representative organisms is shown.  Grey bars refer to all protein residues, while green ones correspond to residues predicted to be in structured, highly stable regions, according to the AlphaFold2 score.
Out of the entire human proteome, the most likely value is to have about 12 sulfur atoms for each protein.

Sulfur atoms belonging to the disordered regions will be characterized by higher movements. Removing them from the statistical analysis conducted on the whole human proteome, the distributions are characterized by a mode of 7 (compared to 12 in the previous case in which all residues have been considered).

This preliminary analysis demonstrates that from a biological point of view, the technique may be applied to perform large screenings as most of the proteins in the proteome have less than 10 sulfur atoms in low motile regions.

From a resolution point of view, the lower the number of sulfur atoms for each protein considered, the higher the resolution of the signal obtained from the scattering. However, with only two atoms we would have a degeneration in the orientation of the protein, thus not being able to obtain information about the binding region. Therefore, the ideal case to perform the experiment is a pair of proteins with three sulfur atoms each. In any case, we expect that the presented methodology may be able to provide reliable results also for proteins with a higher number of sulfur atoms.

\subsection{Molecular dynamics simulations to assess typical relative sulfur movements}

Our protocol does not rely on crystallized protein structures and requires performing two measurements of the sample with beams having lower and higher energies than the sulfur K-edge, to exploit the anomalous dispersion. As the latter depends on the relative distances between the sulfur atoms inside the two proteins when measured alone and in complex, we must consider the relative motion between sulfur atoms, which, ideally, should be fixed both between the two measurements and in different samples. 
To test in which regimes this assumption holds, we performed molecular dynamics simulations of a set of twenty proteins extracted from a larger dataset spanning different protein structures, families, and types.  As one can see from Figure~\ref{fig::3}, the number of sulfur atoms ranges from one to about twenty across the considered proteins, and their spatial disposition ranges along the whole protein structure. Note that proteins tend to form sulfur bridges, so it is frequent to find couples of sulfur atoms at distances lower than 3 $\AA$. Such bridges have a local stabilizing effect so that cysteines and methionines belonging to ordered regions tend to have lower fluctuations than other parts of the structure. In particular, evaluating the Root Mean Squared Fluctuation (RMSF) of all protein residues (see Figure~\ref{fig::3}) confirms this trend.  
Finally, to get an estimate of the time scales of sulfur atoms relative motion, we computed the average difference in the distances of the sulfur atoms of the considered dataset in time. Looking at the pico-second time scale, the average variation of relative distances is 0.5 $\AA$, a tenfold smaller than the needed wavelength.

\subsection{Description of the proposed experiment}

\begin{figure}
\centering
\includegraphics[width=0.7\columnwidth]{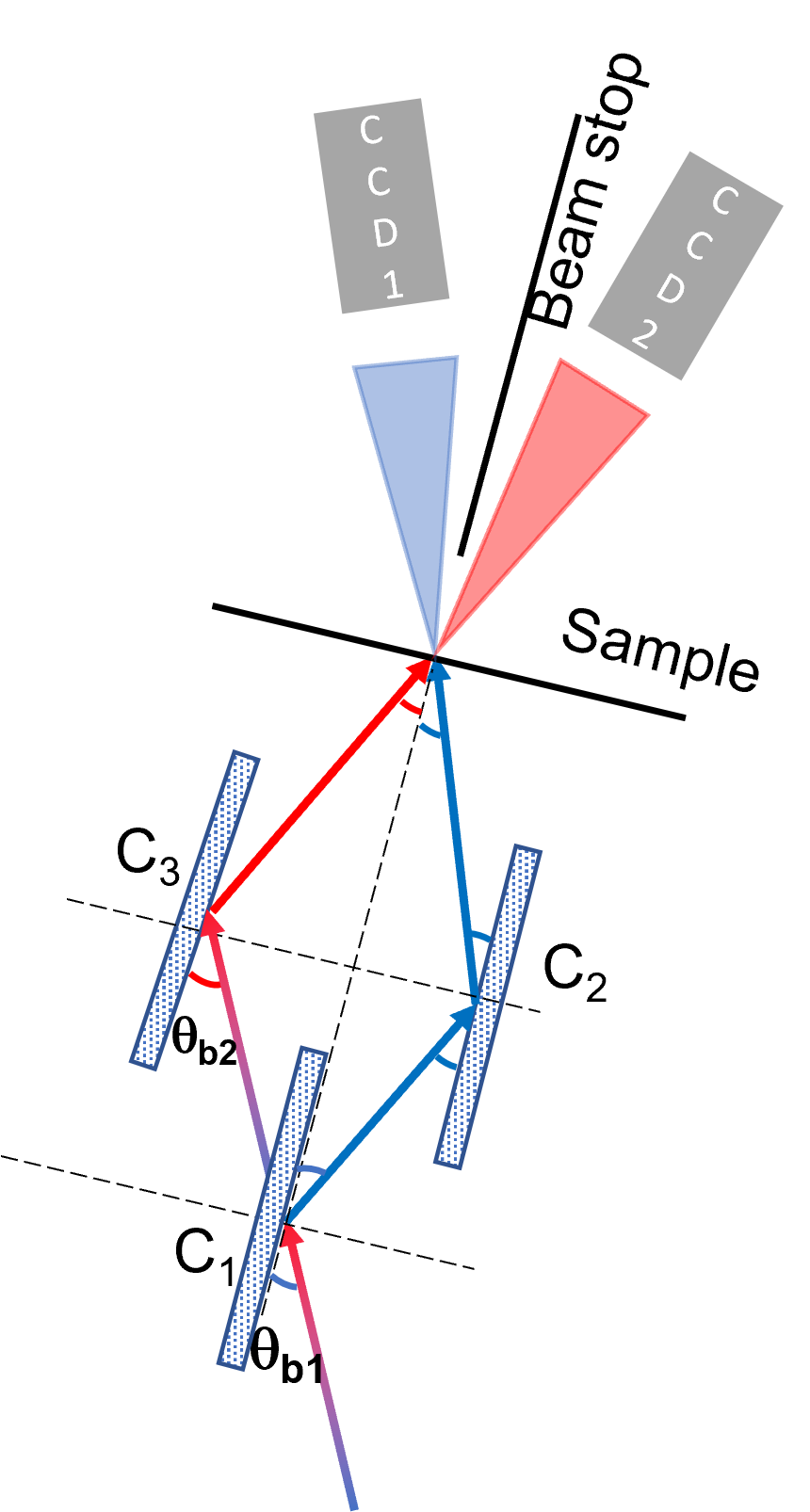}
\caption{\textbf{Sketch of the experimental set-up.} A two-color FEL emission (red-blue arrow) is sent to crystal C1, which is set at the Bragg angle ($\theta_{b1}$) only for the 2.5 KeV photons. The reflected beam (blue arrow) is then intercepted by the C2 crystal which operates at the same Bragg angle, steering it toward the sample. The C3 crystal operated at the Bragg angle for the 2.4 KeV $\theta_{b2}$ emission reflects toward the sample only that color. Two X-ray cameras are placed after the sample to measure the scattering from the proteins. Colors cross talk is avoided by using a metal foil.    }
\label{fig::4}
\end{figure}

 Recent advances in X-ray sources have made available the possibility to realize multiple colors, at about mJ, emissions. Depending on the spectral range, few techniques have been implemented that permit two colors well separated in energy, in time, or both \cite{Allaria2013,serkez_two_2019}. Of particular interest for this work, the latter option is also available around the sulfur K-edge where two pulses separated by up to 2 ps can be emitted with a temporal duration of a few fs. Using the experimental set-up sketch represented in Figure~\ref{fig::4}, the two colors can be separated and recombined at the sample position. Hypothesizing to use a Si111 set of crystals, C1 can be oriented to reflect only the 2.5 KeV beam at an angle $\theta_{b1}$ of 4.53 degrees. A second crystal (C2) can subsequently be used to steer such a beam toward the sample position. A third crystal (C3) placed to intercept the beam transmitted by C1, can be set to reflect the 2.4 KeV at $\theta_{b2}$ equal to 
 4.72. The setup can be designed so that the optical paths of the beams after the point of impact on C1 can compensate for the delay difference between the two X-ray pulses. However, this is not a critical parameter since delays of a few ps can be tolerated. Indeed, the re-orientational dynamics of proteins are in the range of nanoseconds~\cite{Bashardanesh2019}, and pulses separated by less than this value will essentially see frozen sulfur relative distances. 
Two CCD cameras, centered on the transmitted beams will record the scattered photons from the two used pulses. By subtracting the two recorded signals, we will get the scattering contribution of sulfur atoms. Since scattering is expected to extend for more than 15 degrees, a beam stop between the two CCDs will be installed to ensure that each CCD will detect mainly one color. The presence of a residual overlap of the other signal in each camera, i.e. errors in the beam stop component, is expected to slightly reduce the contrast between the two signals. However, we expect this effect not to impact the overall capability of discriminating between bound and unbound proteins.

\section{Discussion}

The determination of protein 3D structure and protein-protein interactions are ranked among the 125 open problems of the century \cite{Kennedy2005}. After a more than fifty-year-long race, the recent development of Artificial Intelligence (AI) based platforms is allowing for the rapid computational prediction of protein structures starting from their amino acid sequence \cite{jumper2021highly}. 
AI-based software, like the novel AlphaFold2 neural network~ \cite{jumper2021highly} is now able to predict protein structures with atomic accuracy starting from their amino acid sequences even in cases in which no similar structures are known, thus overcoming the need to use homology modeling templates, and presents an accuracy competitive with experimental data in most of the cases as assessed in the last CASP competition~\cite{Elofsson2023}. 
Indeed, recently, the entire human proteome was predicted and made available to the scientific community. This achievement paves the way to the subsequent still-open problem, i.e. determining protein-protein interactomes.
Computational advances in this contest are also being made \cite{li2017sprint, pitre2006pipe, martin2005predicting, shen2007predicting, zhang2011adaptive, Milanetti2021, Desantis2022, Grassmann2023}. The abundance of predicted interactions vouches for rapid and efficient experimental techniques able to both validate and guide these predictions. 

Here, we suggest using sulfur atoms anomalous scattering as a way to rapidly detect binding between couples of proteins. 
In particular, analyzing the composition of the proteomes of four organisms widely used in protein-protein interaction studies, we found that proteins contain on average seven sulfur atoms preferentially located on low-motile regions of the protein structure, whose relative distances remain stable on the picosecond timescales. 
A minimal model for the photon scattering of a set of sulfur atoms predicts a difference between the signal of unbound and bound proteins that depends on the relative distances between all the couples of sulfur atoms.  Thus the signal coming from intermolecular sulfur couples permits for a rapid detection of the binding.

Leveraging on the carried-out calculations, we finally proposed an experimental setup to actually measure the interactions. We expect that the time scale for a single measure will be in the second(s) range, which would allow for a high-throughput scanning of molecule interactions.  
If confirmed, we envisage that our proposed technique will be determinant in addressing the future challenge of protein interactome determination as it will permit us to tackle the scanning of the tens of millions of possible couples of dimeric complexes that form the interactomes of complex organisms like humans.

\section{Methods}

\subsection{Protein dataset} 
We consider the dataset proposed by Hensen \textit{et al.}~\cite{Hensen2012}, where a collection of 112 representative proteins for each family was reported.
From this initial set, we selected the 20 proteins, having (i) longer sequences and  (ii) no missing or incomplete residues (see Di Rienzo \textit{et al.}~\cite{di2021characterizing} for further details). For each protein, a molecular dynamics simulation with explicit solvent was performed.

\subsection{Molecular dynamics simulation}

The following protocol was used for each of the 20 simulations.
We used Gromacs 2020~\cite{gromacs} and built the system topology using the CHARMM-27 force field~\cite{charmm}.
The protein was placed in a dodecahedron simulative box, with periodic boundary conditions, filled with TIP3P water molecules~\cite{Jorgensen1983}. We checked that each atom of the protein was at least at a distance of $1.1 \, \mathrm{nm}$ from the box borders.
The system was then minimized with the steepest descent algorithm. Next, a relaxation of water molecules and thermalization of the system was run in NVT and NPT environments each for $0.1 \, \mathrm{ns}$ at $2 \, \mathrm{fs}$ time-step.
The temperature was kept constant at $300 \, \mathrm{K}$ with the v-rescale algorithm \cite{vrescale}; the final pressure was fixed at $1 \, \mathrm{bar}$ with the Parrinello-Rahman algorithm \cite{parrinello}. 
LINCS algorithm \cite{lincs} was used to constrain h-bonds.
A cut-off of $12$ $\AA$ was imposed for the evaluation of short-range non-bonded interactions and the Particle Mesh Ewald method~\cite{Cheatham1995} for the long-range electrostatic interactions. 
Finally, we performed $60 \, \mathrm{ns}$ of molecular dynamics with a time step of $2 \, \mathrm{fs}$, saving configurations every $2 \, \mathrm{ps}$. We considered the last $20 \, \mathrm{ns}$ (10000 frames) for the analysis.

\subsection{Code and data availability}
All relevant data are within the Main Text. Codes can be made available upon reasonable request to the authors.

\section*{Acknowledgements}
The authors acknowledge support by the European Research Council through its Synergy grant program, project ASTRA (grant agreement No 855923), and by the European Innovation Council through its Pathfinder Open Programme, project ivBM-4PAP (grant agreement No 101098989).



\end{document}